%% file: distprob.tex
\documentclass{article}
\usepackage{graphicx}

\begin{document}

\title{Approximate Conditional Distributions of Distances Between Nodes in a
Two-Dimensional Sensor Network}

\author{Rodrigo~S.~C.~Le\~ao\\
Valmir~C.~Barbosa\thanks{Corresponding author (valmir@cos.ufrj.br).}\\
\\
Universidade Federal do Rio de Janeiro\\
Programa de Engenharia de Sistemas e Computa\c c\~ao, COPPE\\
Caixa Postal 68511\\
21941-972 Rio de Janeiro - RJ, Brazil}

\date{}

\maketitle

\begin{abstract}
When we represent a network of sensors in Euclidean space by a graph, there are
two distances between any two nodes that we may consider. One of them is the
Euclidean distance. The other is the distance between the two nodes in the
graph, defined to be the number of edges on a shortest path between them. In
this paper, we consider a network of sensors placed uniformly at random in a
two-dimensional region and study two conditional distributions related to these
distances. The first is the probability distribution of distances in the graph,
conditioned on Euclidean distances; the other is the probability density
function associated with Euclidean distances, conditioned on distances in the
graph. We study these distributions both analytically (when feasible) and by
means of simulations. To the best of our knowledge, our results constitute the
first of their kind and open up the possibility of discovering improved
solutions to certain sensor-network problems, as for example sensor
localization.

\end{abstract}

\section{Introduction}

We consider a network of $n$ sensors, each one placed at a fixed position in
two-dimensional space and capable of communicating with another sensor if and
only if the Euclidean distance between the two is at most $R$, for some constant
radius $R>0$. If $\delta_{ij}$ denotes this distance for sensors $i$ and $j$,
then a graph representation of the network can be obtained by letting each
sensor be a node and creating an edge between any two distinct nodes $i$ and $j$
such that $\delta_{ij}\le R$. Such a representation is, aside from a scale
factor, equivalent to a unit disk graph \cite{disk}.

Often $n$ is a very large integer and the network is essentially unstructured,
in the sense that the sensors' positions, although fixed, are generally unknown.
In domains for which this holds, generalizing the graph representation in such a
way that each node's position is given by random variables becomes a crucial
step, since it opens the way to the investigation of relevant distributions
related to all networks that result from the same deployment process. Such a
generalization, which can be done for any number of dimensions, is known as a
random geometric graph \cite{penrose}. Similarly to the random graphs of
Erd\H{o}s and R\'{e}nyi \cite{erdos} and related structures \cite{nsw01}, many
important properties of random geometric graphs are known, including some
related to connectivity and the appearance of the giant component 
\cite{appel_connect,appel_maxdegree,appel_mindegree} and others more closely
related to applications \cite{percolation,gupta,mcdiarmid}.

One curious aspect of random geometric graphs is that, if nodes are positioned
uniformly at random, the expected Euclidean distance between any two nodes is a
constant in the limit of very large $n$, depending only on the number of
dimensions (two, in our case) \cite{box}. In this case, distance-dependent
analyses must necessarily couple the Euclidean distance with some other type of
distance between nodes. The natural candidate is the standard graph-theoretic
distance between two nodes, given by the number of edges on a shortest path
between them \cite{bondy}. For nodes $i$ and $j$, this distance is henceforth
denoted by $d_{ij}$ and referred to simply as the distance between $i$ and $j$.

Given $i$ and $j$, the Euclidean distance $\delta_{ij}$ and the distance
$d_{ij}$ between the two nodes are not independent of each other, but rather
interrelate in a complex way. Our goal in this paper is to explore the
relationship between the two when all sensors are positioned uniformly at random
in a given two-dimensional region. Specifically, for $i$ and $j$ two distinct
nodes chosen at random, we study the probability that $d_{ij}=d$ for some
integer $d>0$, given that $\delta_{ij}=\delta$ for some real number
$\delta\ge 0$. Similarly, we also study the probability density associated with
$\delta_{ij}=\delta$ when $d_{ij}=d$. Our study is analytical whenever feasible,
but is also computational throughout. Depending on the value of $d$, we are in a
few cases capable of providing exact closed-form expressions, but in general
what we give are approximations, either derived mathematically or inferred from
simulation data exclusively.

We remark, before proceeding, that we perceive the study of distance-related
distributions for random geometric graphs as having great applicability in the
field of sensor networks, particularly in domains in which it is important for
each sensor to have a good estimate of its location. In fact, of all possible
applications that we normally envisage for sensor networks \cite{instrumenting},
network localization is crucial in all cases that require the sensed data to be
tagged with reliable indications of where the data come from; it has also been
shown to be important even for routing purposes \cite{routing}. So, although we
do not dwell on the issue of network localization anywhere else in the paper, we
now digress momentarily to clarify what we think the impact of distance-related
distributions may be.

The problem of network localization has been tackled from a variety of
perspectives, including rigidity-theoretic studies \cite{rigid,theory},
approaches that are primarily algorithmic, either centralized
\cite{convex,local_connect} or distributed
\cite{range_free,dv_based,global_coord,robust_noisy}, and others that generalize
on our assumptions by taking advantage of sensor mobility
\cite{local_mobile,robust_noisy} or uneven radii \cite{mds}. In general one
assumes the existence of some anchor sensors (regularly placed \cite{gps-less}
or otherwise), for which positions are known precisely, and then the problem
becomes reduced to the problem of providing, for each of the other sensors, the
Euclidean distances that separate it from three of the anchors (its tripolar
coordinates with respect to those anchors, from which the sensor's position can
be easily calculated \cite{wolfram}).

Finding a sensor's Euclidean distance to an anchor is not simple, though.
Sometimes signal propagation is used for direct or indirect measurement
\cite{radar,spoton,cricket,robust_range,omnet,aps}, but there are approaches
that rely on no such techniques \cite{amorph,gps-less,dv_based}. The latter
include one of the most successful distributed approaches \cite{dv_based}, which
nonetheless suffers from increasing lack of accuracy as sparsity or irregularity
in sensor positioning become more pronounced. The algorithm of \cite{dv_based}
assumes, for each anchor $i$, that each edge on any shortest path to $i$ is
equivalent to a fixed Euclidean distance, which is estimated by $i$ in
communication with the other anchors and by simple proportionality can be used
by any node to infer its Euclidean distance to $i$. We believe that knowledge of
distance-related distributions has an important role to play in replacing this
assumption and perhaps dispelling the algorithm's difficulties in the less
favorable circumstances alluded to above.

We proceed in the following manner. In Section~2 we give some notation and
establish the overall approach to be followed when pursuing the analytical
characterization of distance-related distributions. Then in Sections~3 through
5 we present the mathematical analysis of the $d=1$ through $d=3$ cases. We
continue in Section~6 with computational results related to $d\ge 1$ and close
in Section~7 with some discussion and concluding remarks.
 
\section{Overall approach}

Let $i$ and $j$ be two distinct, randomly chosen nodes. For $d>0$ an integer and
$\delta\ge 0$ a real number, we use $P_\delta(d)$ to denote the probability,
conditioned on $\delta_{ij}=\delta$, that $d_{ij}=d$. Likewise, we use
$p_d(\delta)$ to denote the probability density, conditioned on $d_{ij}=d$,
associated with $\delta_{ij}=\delta$. These two quantities relate to each other
in the standard way of combining integer and continuous random variables
\cite{trivedi}.

If we assume that $P_\delta(d)$ is known for all applicable values of $d$ and
$\delta$, then it follows from Bayes' theorem that
\begin{equation}
p_d(\delta)
=\frac{P_\delta(d)p(\delta)}{P(d)},
\end{equation}
where $p(\delta)$ is the unconditional probability density associated with the
occurrence of an Euclidean distance of $\delta$ separating two randomly chosen
nodes and $P(d)$ is the unconditional probability that the distance between them
is $d$. Clearly, $P(d)=\int_{r=0}^{dR}P_r(d)p(r)\mathrm{d}r$, since $P_r(d)=0$
for $r>dR$. Moreover, $p(r)$ is proportional to the circumference of a
radius-$r$ circle, $2\pi r$, which yields
\begin{equation}
p_d(\delta)=\frac{P_\delta(d)\delta}{\int_{r=0}^{dR}P_r(d)r\mathrm{d}r}.
\label{eq:pfromP}
\end{equation}

In view of Equation (\ref{eq:pfromP}), our approach henceforth is to concentrate
on calculating $P_\delta(d)$ for all appropriate values of $d$ and $\delta$,
and then to use the equation to obtain $p_d(\delta)$. In order to calculate
$P_\delta(d)$, we fix two nodes $a$ and $b$ such that $\delta_{ab}=\delta$ and
proceed by analyzing how the two radius-$R$ circles (the one centered at $a$ and
the one at $b$) relate to each other. While doing so, we assume that the
two-dimensional region containing the graph has unit area, so that the area of
any of its sub-regions automatically gives the probability that it contains a
randomly chosen node. We assume further that all border effects can be safely
ignored (but see Section~6 for the computational setup that justifies this).

\section{The distance-1 and distance-2 cases}

The case of $d=1$ is straightforward, since $d_{ab}=d$ if and only if
$\delta\le R$. Consequently,
\begin{equation}
P_\delta(1)=\cases{
1,&if $\delta\le R$;\cr
0,&otherwise
}
\end{equation}
and, by Equation (\ref{eq:pfromP}),
\begin{equation}
p_1(\delta)=\cases{
2\delta/R^2,&if $\delta\le R$;\cr
0,&otherwise.
}
\end{equation}

For $d=2$, we have $d_{ab}=d$ if and only if $\delta >R$ and at least one node
$k$ exists, with $k\notin\{a,b\}$, such that $\delta_{ak}\le R$ and
$\delta_{bk}\le R$. The probability that this holds for a randomly chose $k$ is
given by the intersection area of the radius-$R$ circles centered at $a$ and
$b$, here denoted by $\rho_\delta$. From \cite{wolfram}, we have
\begin{equation}
\rho_\delta=\cases{
2R^2\cos^{-1}\left(\delta/2R\right)
-\delta\sqrt{R^2-\delta^2/4},&if $\delta\le 2R$;\cr
0,&otherwise.
}
\label{eq:intersection}
\end{equation}
Because any node that is not $a$ or $b$ may, independently, belong to such
intersection, we have
\begin{equation}
P_\delta(2)=\cases{
1-(1-\rho_\delta)^{n-2},&if $\delta>R$;\cr
0,&otherwise.
}
\end{equation}
As for $p_2(\delta)$, it is as given by Equation (\ref{eq:pfromP}), equaling $0$
if $\delta\le R$ or $\delta>2R$ (we remark that a closed-form expression is
obtainable also in this case, but it is too cumbersome and is for this reason
omitted).

\section{The distance-3 case: exact basis}

The $d=3$ case is substantially more complex than its predecessors in Section~3.
We begin by noting that $d_{ab}=d$ if and only if the following three conditions
hold:
\begin{itemize}
\item[C1.] $\delta>R$.
\item[C2.] No node $i$ exists such that both $\delta_{ai}\le R$ and
$\delta_{bi}\le R$.
\item[C3.] At least one node $k\notin\{a,b\}$ exists, and for this $k$ at least
one node $\ell\notin\{a,b,k\}$, such that $\delta_{ak}\le R$,
$\delta_{k\ell}\le R$, $\delta_{b\ell}\le R$, $\delta_{a\ell}>R$, and finally
$\delta_{bk}>R$.
\end{itemize}
For each fixed $k$ and $\ell$ in Condition~C3, these three conditions result
from the requirement that nodes $a$, $k$, $\ell$, and $b$, in this order,
constitute a shortest path from $a$ to $b$.

\begin{figure}[t]
\centering
\begin{tabular}{c}
\includegraphics[scale=0.65]{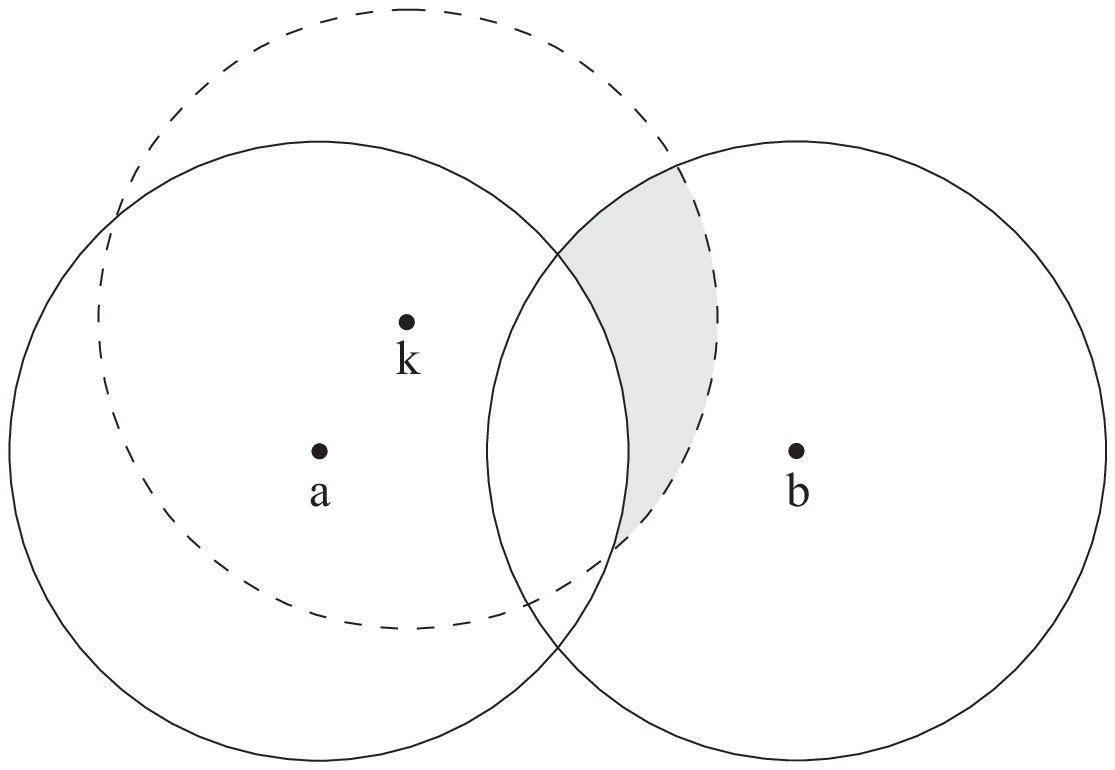}\\
{\small (a)}\\
\includegraphics[scale=0.65]{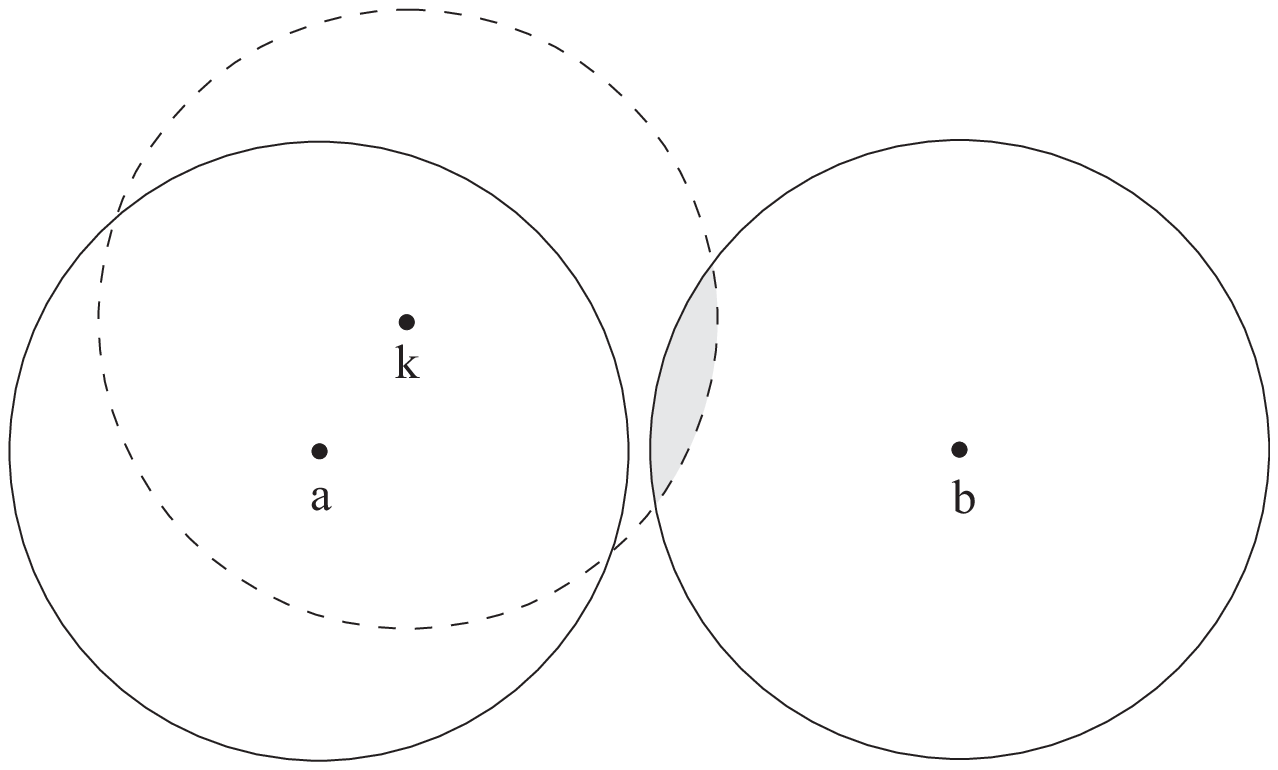}\\
{\small (b)}
\end{tabular}
\caption{Regions (shown in shades) whose areas yield the value of
$\sigma_\delta^k$ for $R<\delta\le 2R$ (a) and $2R<\delta\le 3R$ (b).}
\label{fig:2or3circles}
\end{figure}

If we fix some node $k\notin\{a,b\}$ for which $\delta_{ak}\le R$ and
$\delta_{bk}>R$, the probability that Condition~C3 is satisfied by $k$ and a
randomly chosen $\ell$ is a function of intersection areas of circles that
varies from case to case, depending on the value of $\delta$. There are two
cases to be considered, as illustrated in Figure~\ref{fig:2or3circles}. In the
first case, illustrated in part (a) of the figure, $R<\delta\le 2R$ and node
$\ell$ is to be found in the intersection of the radius-$R$ circles centered at
$b$ and $k$, provided it is not also in the radius-$R$ circle centered at $a$.
The intersection area of interest results from computing the intersection area
of two circles (those centered at $b$ and $k$) and subtracting from it the
intersection area of three circles (those centered at $a$, $b$, and $k$). The
former of these intersection areas is given as in
Equation~(\ref{eq:intersection}), with $\delta_{bk}$ substituting for $\delta$;
as for the latter, closed-form expressions also exist, as given in
\cite{3circles}. The second case, shown in part (b) of
Figure~\ref{fig:2or3circles}, is that of $2R<\delta\le 3R$, and then the
intersection area of interest is the one of the circles centered at $b$ and $k$.
Regardless of which case it is, we use $\sigma_\delta^k$ to denote the resulting
area. Thus, the probability that at least one $\ell$ exists for fixed $k$ is
$1-(1-\sigma_\delta^k)^{n-3}$.

Now let $P'_\delta(3)$ be the probability that a randomly chosen $k$ satisfies
Condition~C3. Let also $K_\delta$ be the region inside which such a node can be
found with nonzero probability. If $x_k$ and $y_k$ are the Cartesian coordinates
of node $k$, then each possible location of $k$ inside $K_\delta$ contributes to
$P'_\delta(3)$ the infinitesimal probability
$[1-(1-\sigma_\delta^k)^{n-3}]\mathrm{d}x_k\mathrm{d}y_k$. It follows that
\begin{equation}
P'_\delta(3)=\int_{k\in K_\delta}
[1-(1-\sigma_\delta^k)^{n-3}]\mathrm{d}x_k\mathrm{d}y_k.
\end{equation}

\begin{figure}[p]
\centering
\begin{tabular}{c}
\includegraphics[scale=0.65]{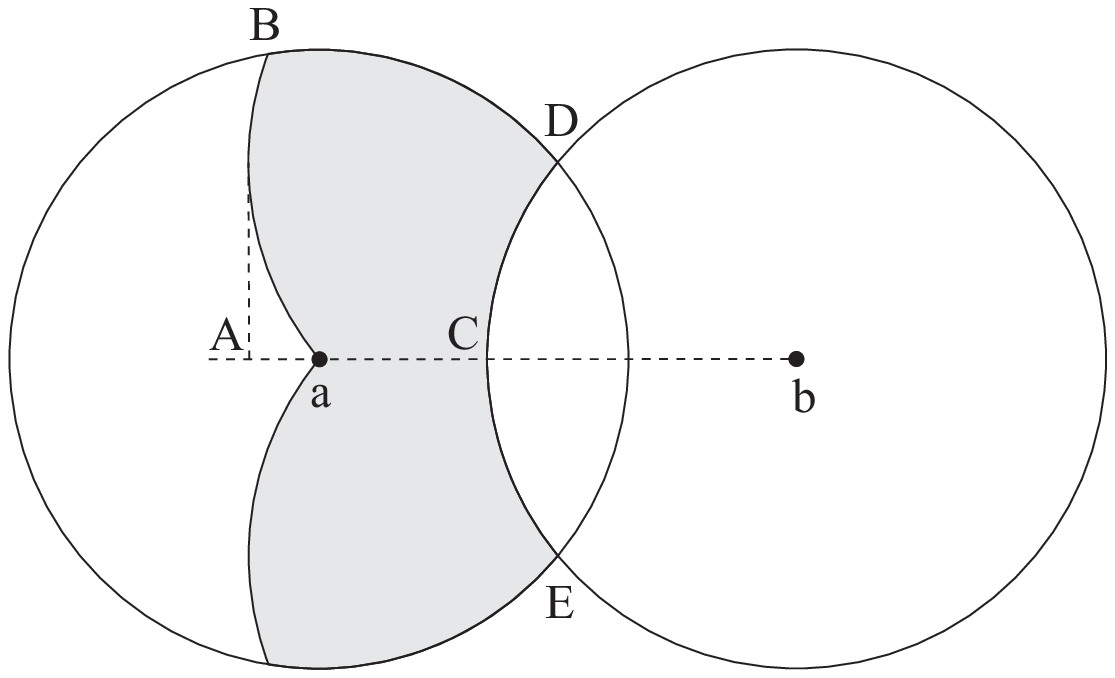}\\
{\small (a)}\\
\includegraphics[scale=0.65]{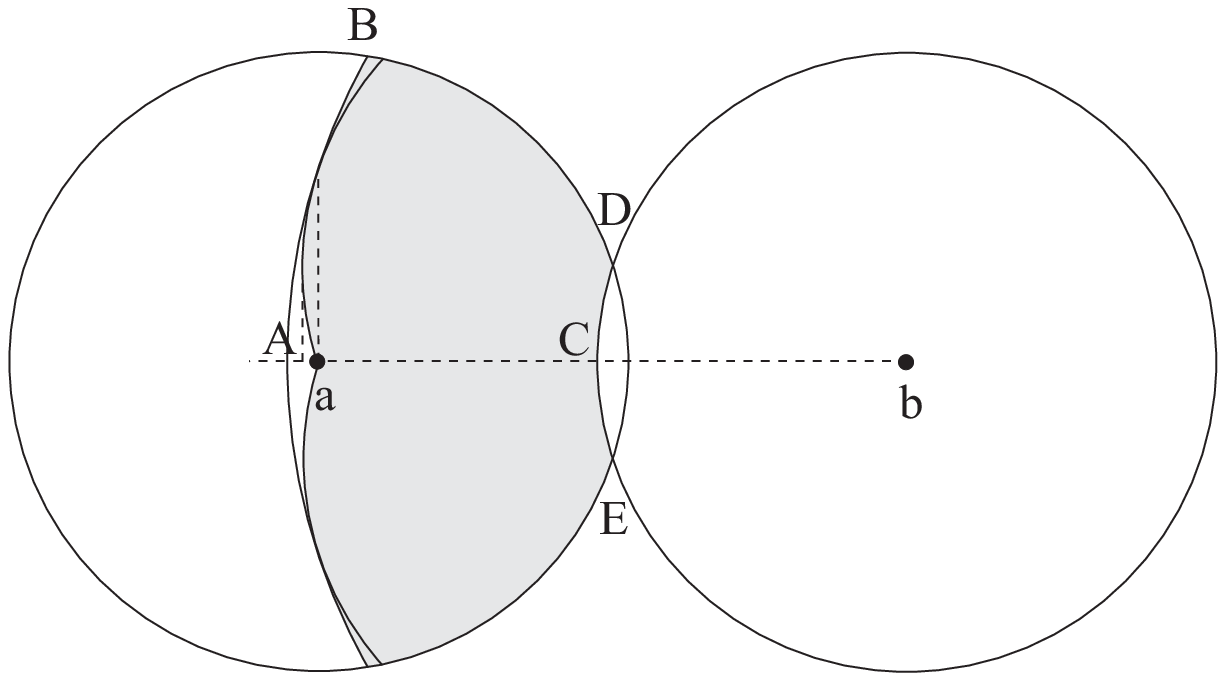}\\
{\small (b)}\\
\includegraphics[scale=0.65]{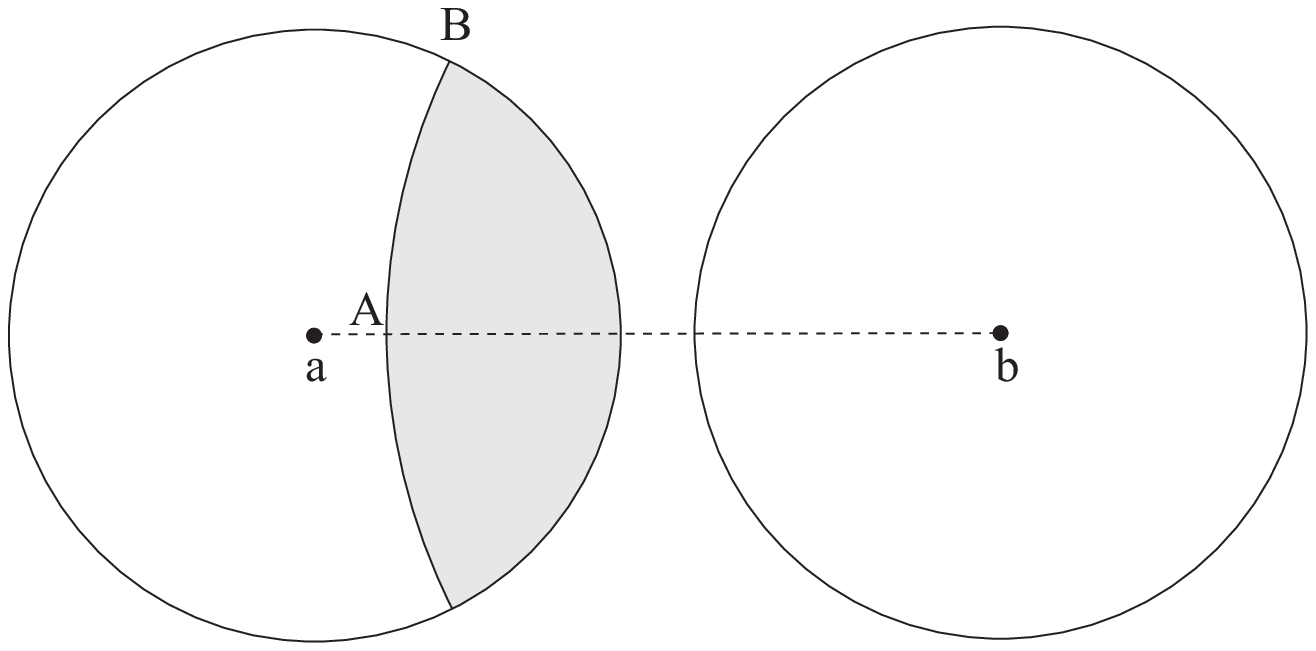}\\
{\small (c)}
\end{tabular}
\caption{Regions (shown in shades) where node $k$ can be found with nonzero
probability for $R<\delta\le R\sqrt{3}$ (a), $R\sqrt{3}<\delta\le 2R$ (b), and
$2R<\delta\le 3R$ (c).}
\label{fig:Kdelta}
\end{figure}

There are three possibilities for the region $K_\delta$, shown in parts (a)
through (c) of Figure~\ref{fig:Kdelta} as shaded regions, respectively for
$R<\delta\le R\sqrt{3}$, $R\sqrt{3}<\delta\le 2R$, and $2R<\delta\le 3R$. The
shaded region in part (a) is delimited by four radius-$R$ circles, the ones
centered at nodes $a$ (above and below) and $b$ (on the right) and the ones
centered at points $D$ and $E$ (on the left). As $\delta$ gets increased beyond
$R\sqrt{3}$---and, at the threshold, point $D$ becomes collinear with point $B$
and node $b$---we move into part (b) of the figure, where the shaded region is
now delimited on the left either by the radius-$R$ circles centered at $D$ and
$E$ or by the radius-$2R$ circle centered at $b$, depending on the point of
common tangent between each of the radius-$R$ circles and the radius-$2R$
circle. The next threshold leads $\delta$ beyond $2R$, and in part (c) of the
figure the shaded region is delimited on the left by the radius-$2R$ circle
centered at $b$, on the right by the radius-$R$ circle centered at $a$.

Figure~\ref{fig:Kdelta} is also useful in helping us obtain a more operational
version of the expression for $P'_\delta(3)$, to be used in Section~6. First we
establish a Cartesian coordinate system by placing its origin at node $a$ and
making the positive abscissa axis go through node $b$. In this system, the
shaded regions in all of parts (a) through (c) of the figure are symmetrical
with respect to the abscissa axis. If for each value of $x_k$ we let
$y_k^-(x_k)$ and $y_k^+(x_k)$ be, respectively, the minimum and maximum $y_k$
values in the upper half of the shaded region for the value of $\delta$ at hand,
then
\begin{equation}
P'_\delta(3)=2\int_{x_k=x_k^-}^{x_k^+}\int_{y_k=y_k^-(x_k)}^{y_k^+(x_k)}
[1-(1-\sigma_\delta^k)^{n-3}]\mathrm{d}x_k\mathrm{d}y_k,
\end{equation}
where $x_k^-$ and $x_k^+$ bound the possible values of $x_k$ for the given
$\delta$.

All pertinent values of $x_k^-$ and $x_k^+$, as well as of $y_k^-(x_k)$ and
$y_k^+(x_k)$, are given in Table~\ref{tab:limits}, where $\delta^-$ and
$\delta^+$ indicate, respectively, the lower and upper limit for $\delta$ in
each of the three possible cases. This table's entries make reference to the
abscissae of points $A$, $B$, $C$, and $D$ (respectively $x_A$, $x_B$, $x_C$,
and $x_D$) and to the ordinate of point $D$ ($y_D$). These are given in
Table~\ref{tab:nodes}.

\begin{table}[p]
\centering
{\renewcommand{\arraystretch}{1.5}
\caption{Cartesian coordinates delimiting the upper halves of shaded regions in
Figure~\ref{fig:Kdelta}.}
\label{tab:limits}
{\small\input{tab1}}
}
\end{table}

\begin{table}[p]
\centering
{\renewcommand{\arraystretch}{1.5}
\caption{Cartesian coordinates used in Table~\ref{tab:limits}.}
\label{tab:nodes}
{\small\input{tab2}}
}
\end{table}

\section{The distance-3 case: approximate extension}

Obtaining $P_\delta(3)$ from $P'_\delta(3)$ requires that we fulfill the
remaining requirements set by Conditions~C2 and C3 in Section~4. These are that
no node exists in the intersection of the radius-$R$ circles centered at $a$ and
$b$ and that at least one node $k$ exists with the properties given in
Condition~C3. While the probability of the former requirement is simply
$(1-\rho_\delta)^{n-2}$, expressing the probability of the latter demands that
we make a careful approximation to compensate for the lack of independence of
certain events with respect to one another.

For node $i\notin\{a,b\}$, let $\epsilon_i$ stand for the event that
Condition~C3 does not hold for $k=i$. Let also $Q_\delta(\epsilon_i)$ be the
probability of $\epsilon_i$ and $Q_\delta$ the joint probability of all $n-2$
events.  Clearly, $Q_\delta(\epsilon_i)=1-P'_\delta(3)$ for any $i$ and, for
$\delta>R$, $P_\delta(3)=(1-Q_\delta)(1-\rho_\delta)^{n-2}$. Therefore, if all
the $n-2$ events were independent of one another, we would have
\begin{equation}
Q_\delta=\prod_{i\notin\{a,b\}}Q_\delta(\epsilon_i)=[1-P'_\delta(3)]^{n-2}
\end{equation}
and, consequently,
\begin{equation}
P_\delta(3)=\cases{
\{1-[1-P'_\delta(3)]^{n-2}\}(1-\rho_\delta)^{n-2},&if $\delta>R$;\cr
0,&otherwise.
}
\end{equation} 

However, once we know of a certain node $i$ that Condition~C3 does not hold for
it, immediately we reassess as less likely that the condition holds for nodes in
the Euclidean vicinity of $i$. The $n-2$ events introduced above are then not
unconditionally independent of one another, although we do expect whatever
degree of dependence there is to wane progressively as we move away from node
$i$.

We build on this intuition by postulating the existence of an integer $n'<n-2$
such that the independence of the $n'$ events not only holds but is also
sufficient to determine $P_\delta(3)$ as indicated above, provided the
corresponding $n'$ nodes are picked uniformly at random. But since this is
precisely the way in which, by assumption, sensors are positioned, it suffices
that any $n'$ nodes be selected, yielding
\begin{equation}
P_\delta(3)=\cases{
\{1-[1-P'_\delta(3)]^{n'}\}(1-\rho_\delta)^{n-2},&if $\delta>R$;\cr
0,&otherwise.
}
\end{equation}
Similarly to the previous cases, $p_3(\delta)$ is given by
Equation~(\ref{eq:pfromP}) and equals $0$ if $\delta\le R$ or $\delta>3R$.

It remains, of course, for the value of $n'$ to be discovered if our postulate
is to be validated. We have done this empirically, by means of computer
simulations, as discussed in Section~6.

\section{Computational results}

In this section we present simulation results and, for $d=1,2,3$, contrast them
with the analytic predictions of Sections~3 through 5. The latter are obtained
by numerical integration when a closed-form expression is not available (the
case of $d=3$ also requires simulations for finding $n'$; see below). For $d>3$,
we demonstrate that good approximations by Gaussians can be obtained.

We use $n=1\,000$ and a circular region of unit area, therefore of radius
$\sqrt{1/\pi}\approx 0.564$, for the placement of nodes. Node $a$ is always
placed at the circle's center, which has Cartesian coordinates $(0,0)$, and all
results refer to distances to $a$. Our choice for the value of $R$ depends on
the expected number of neighbors (or connectivity) of a node, which we denote by
$z$ and use as the main parameter. Since $z=\pi R^2n$ for large $n$, choosing
the value of $z$ immediately yields the value of $R$ to be used. We use $z=3\pi$
and $z=5\pi$, which yield, respectively, $R\approx 0.055$ and $R\approx 0.071$.
We note that both values of $z$ are significantly above the phase transition
that gives rise to the giant component, which happens at $z\approx 4.52$
\cite{rgg}.  In all our experiments, then, graphs are connected with high
probability.

For each value of $z$, each simulation result we present is an average over
$10^6$ independent trials. Each trial uses a matrix of accumulators having $n-1$
rows (one for each of the possible distance values) and $1\,000\sqrt{1/\pi}$
columns (one for each of the $0.001$-wide bins into which Euclidean distances
are compartmentalized). A trial consists of: placing $n-1$ nodes uniformly at
random in the circle; computing the Euclidean distance between each node and
node $a$; computing the distances between each node and node $a$ (this is done
with Dijkstra's algorithm \cite{cormen}); updating the accumulator that
corresponds to each node, given its two distances. At the end of each trial, its
contributions to $P_\delta(d)$ and $p_d(\delta)$ are computed, with
$d=1,2,\ldots,n-1$ and $\delta$ ranging through the middle points of all bins.
If $M$ is the matrix of accumulators, then these contributions are given,
respectively, by $M(d,\delta)/\sum_{d'}M(d',\delta)$ and
$M(d,\delta)/0.001\sum_{\delta'}M(d,\delta')$.

The case of $d=3$ requires two additional simulation procedures, one for
determining simulation data for $P'_\delta(3)$, the other to determine $n'$ for
use in obtaining analytic predictions for $P_\delta(3)$. The former of these
fixes node $b$ at coordinates $(\delta,0)$ and performs $10^7$ independent
trials. At each trial, two nodes are generated uniformly at random in the
circle. At the end of all trials, the desired probability is computed as the
fraction of trials that resulted in nodes $k$ and $\ell$ as in Section~4.

The simulation for the determination of $n'$ is conducted for $\delta=2R$ only,
whence $\rho_\delta=0$. This is the value of $\delta$ for which the results from
the simulation above for $P_\delta(3)$ and the analytic prediction for
$1-[1-P'_\delta(3)]^{n-2}$ differ the most (data not shown). Moreover, as we
will see shortly, the value of $n'$ we find using this value of $\delta$ is good
for all other values as well. The simulation is aimed at finding the value of
$Q_\delta$ and proceeds in $10^9$ independent trials. Each trial fixes node $b$
at $(\delta,0)$ and places the remaining $n-2$ nodes in the circle uniformly at
random. The fraction of trials resulting in no node qualifying as the node $k$
of Section~4 is the value of $Q_\delta$. We set $n'$ to be the $m<n-2$ that
minimizes $\vert Q_\delta-[1-P'_\delta(3)]^m\vert$, where $P'_\delta(3)$ refers
to the analytic prediction. Our results are $n'=779$ for $z=3\pi$, $n'=780$ for
$z=5\pi$.

\begin{figure}[p]
\centering
\begin{tabular}{c}
\includegraphics[scale=0.65]{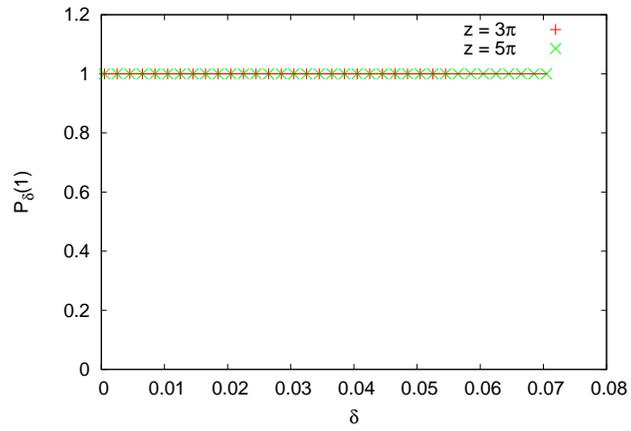}\\
{\small (a)}\\
\includegraphics[scale=0.65]{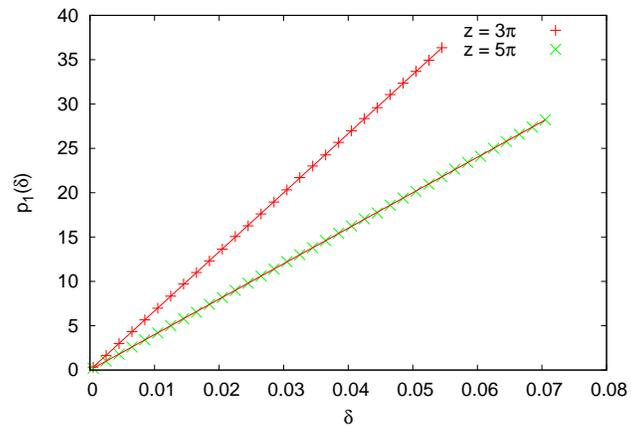}\\
{\small (b)}
\end{tabular}
\caption{$P_\delta(1)$ (a) and $p_1(\delta)$ (b). Solid lines give the analytic
predictions.}
\label{fig:d=1}
\end{figure}

\begin{figure}[p]
\centering
\begin{tabular}{c}
\includegraphics[scale=0.65]{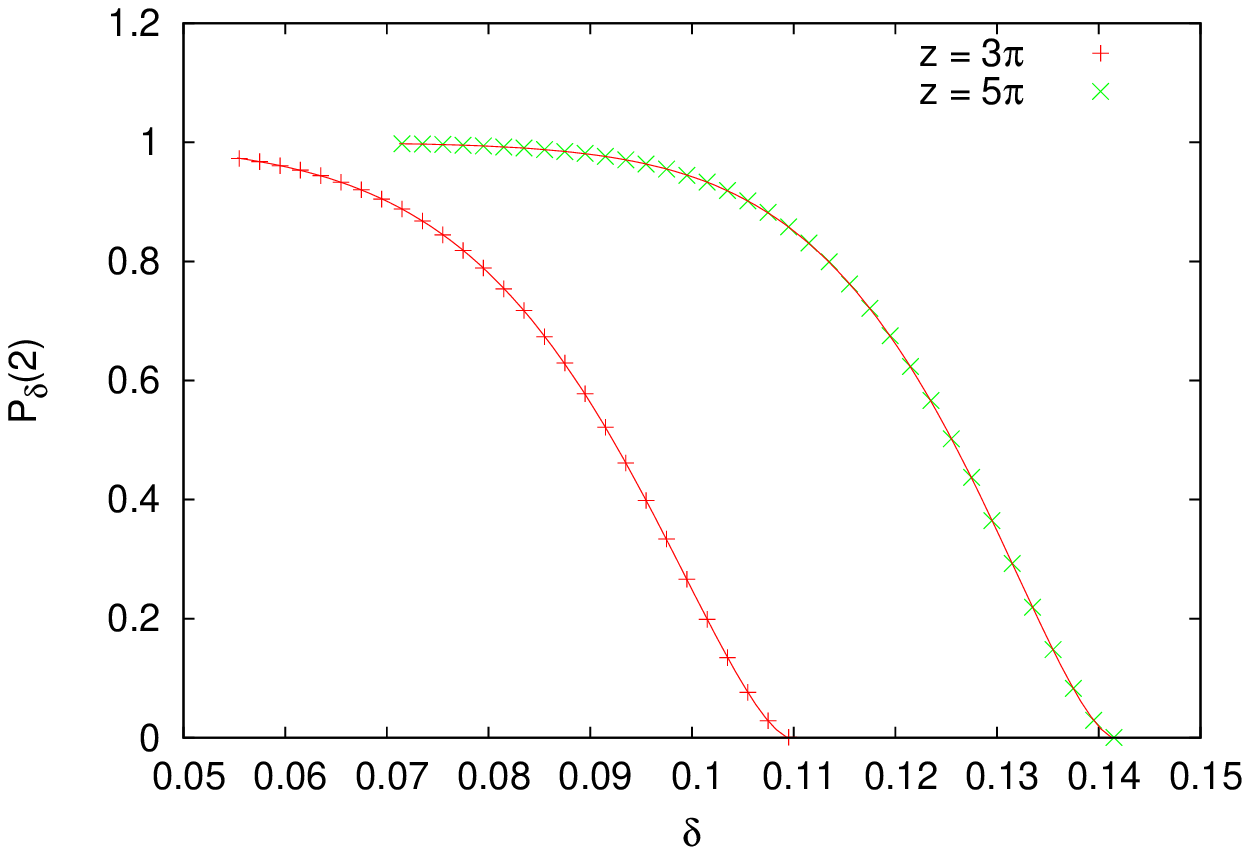}\\
{\small (a)}\\
\includegraphics[scale=0.65]{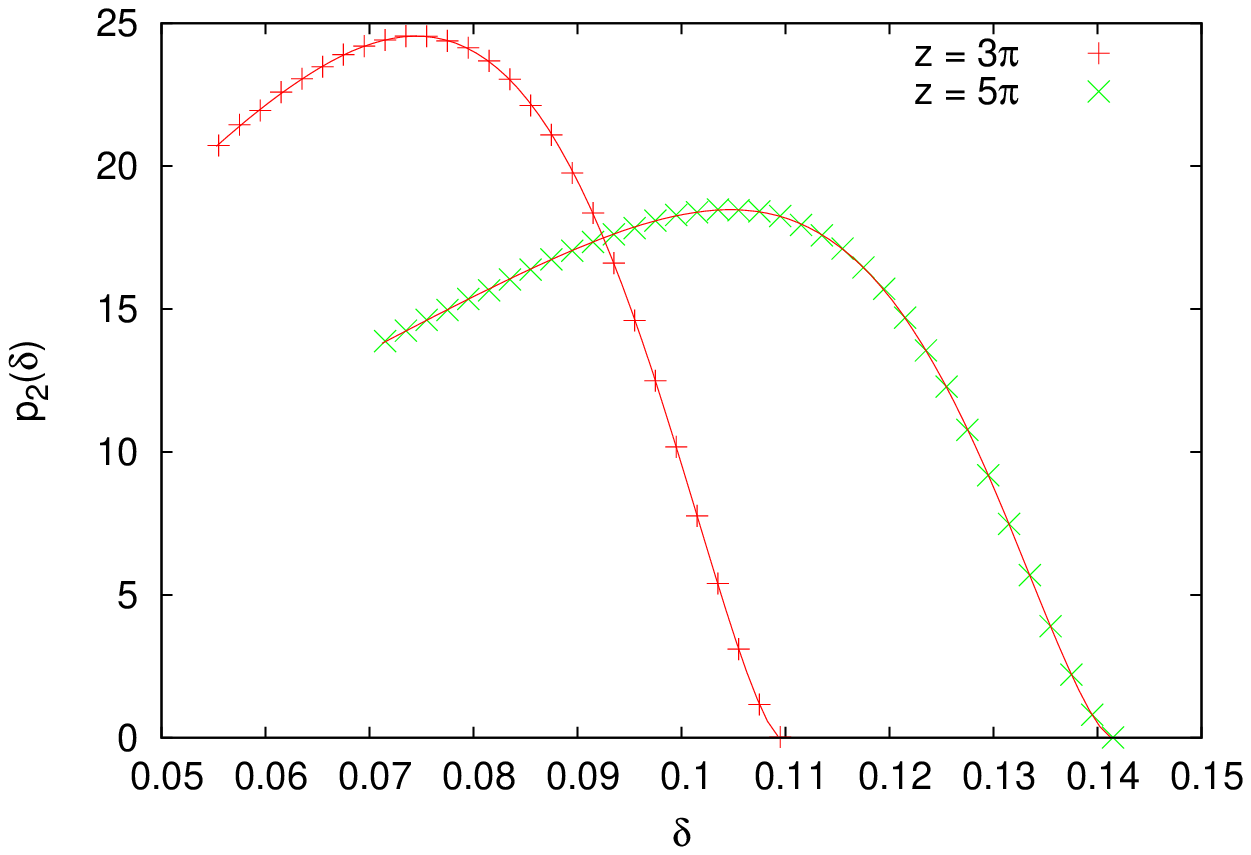}\\
{\small (b)}
\end{tabular}
\caption{$P_\delta(2)$ (a) and $p_2(\delta)$ (b). Solid lines give the analytic
predictions.}
\label{fig:d=2}
\end{figure}

\begin{figure}[t]
\centering
\begin{tabular}{c}
\includegraphics[scale=0.65]{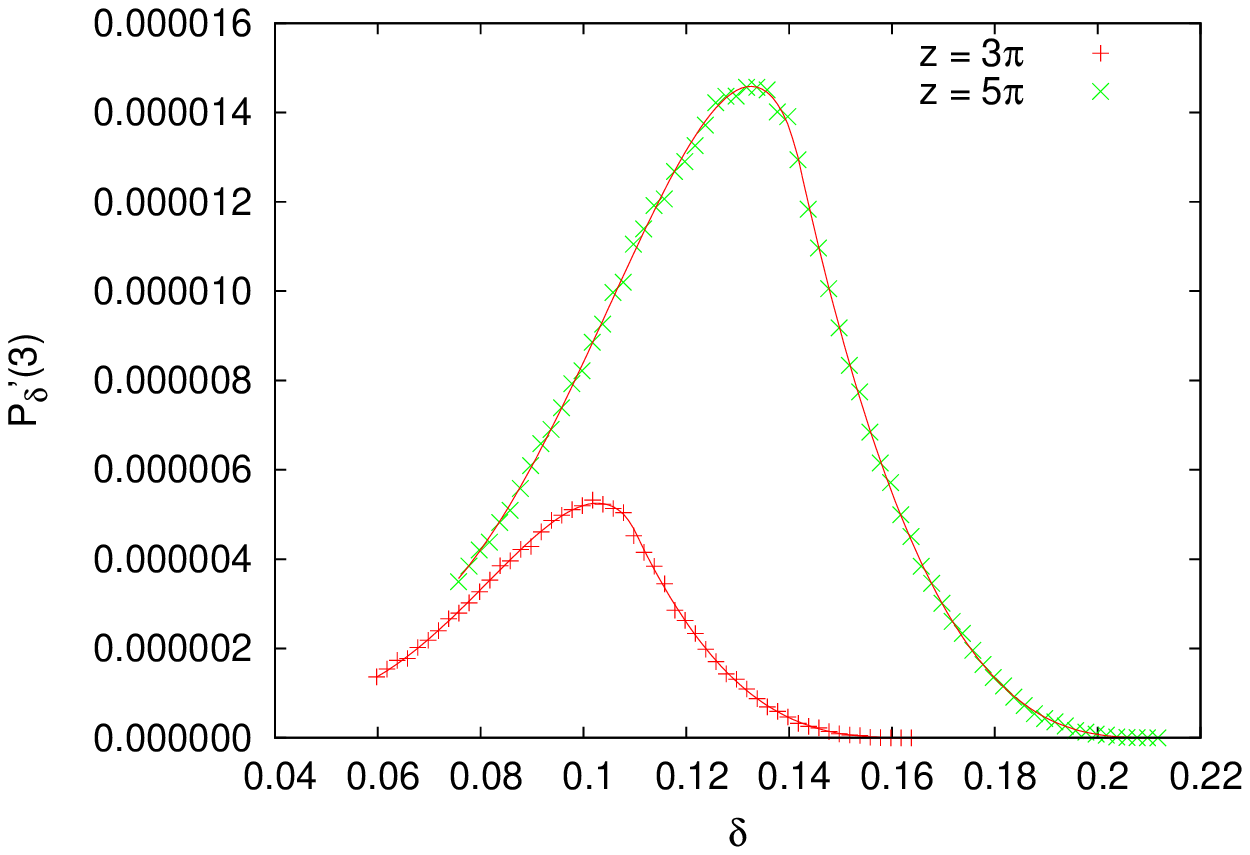}
\end{tabular}
\caption{$P'_\delta(3)$. Solid lines give the analytic predictions.}
\label{fig:P'}
\end{figure}

\begin{figure}[p]
\centering
\begin{tabular}{c}
\includegraphics[scale=0.65]{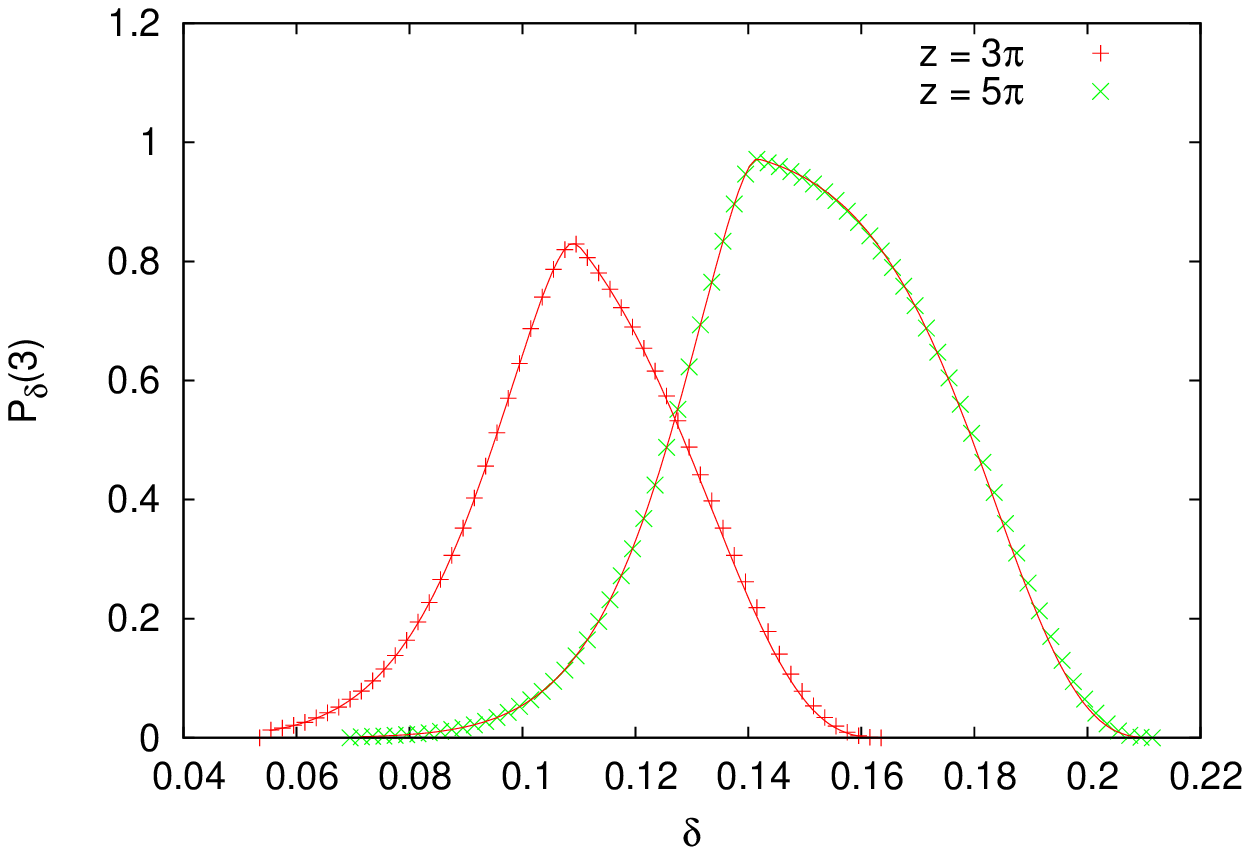}\\
{\small (a)}\\
\includegraphics[scale=0.65]{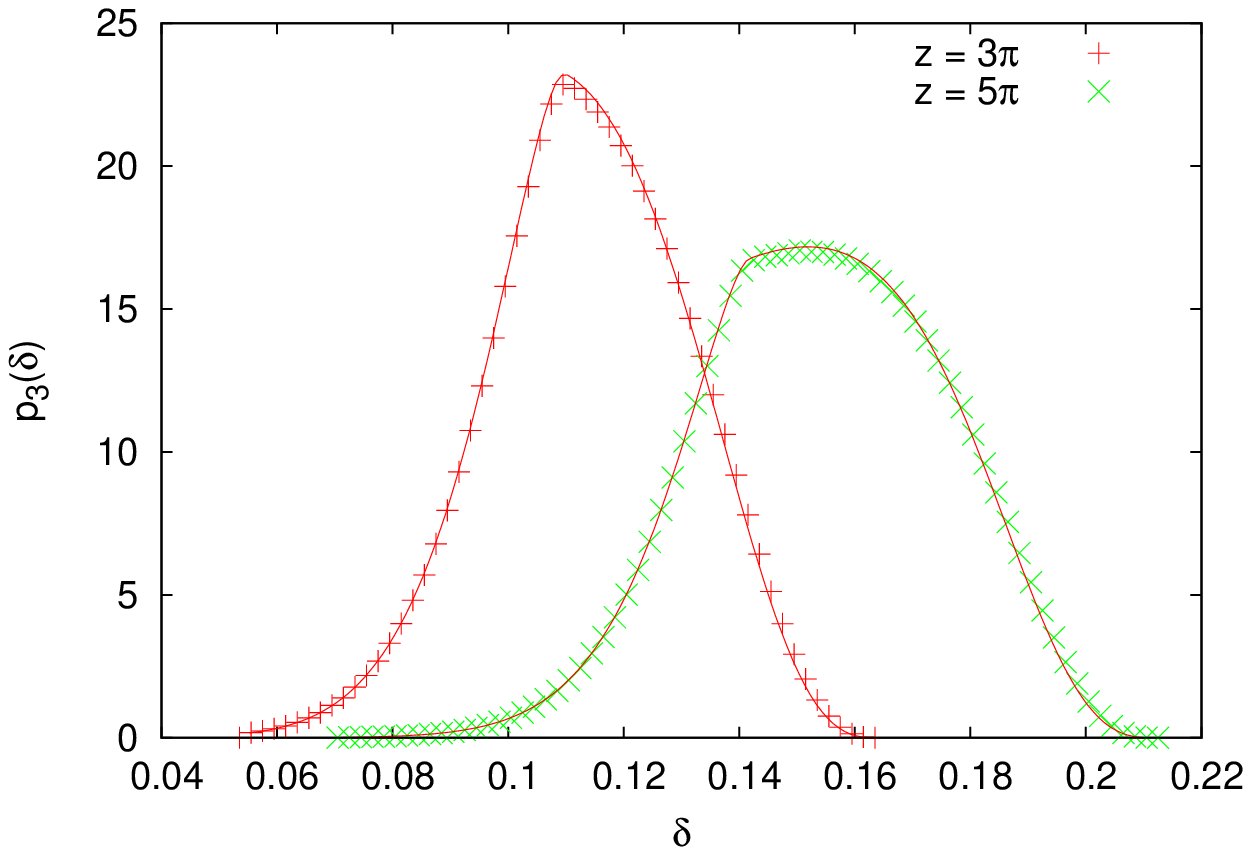}\\
{\small (b)}
\end{tabular}
\caption{$P_\delta(3)$ (a) and $p_3(\delta)$ (b). Solid lines give the analytic
predictions.}
\label{fig:d=3}
\end{figure}

Results for $d=1$ are shown in Figure~\ref{fig:d=1}, for $d=2$ in
Figure~\ref{fig:d=2}, for $d=3$ in Figures~\ref{fig:P'} and \ref{fig:d=3}, and
for $d>3$ in Figure~\ref{fig:d>3}. In all figures, both $P_\delta(d)$ and
$p_d(\delta)$ are plotted against $\delta$, since it seems better to visualize
what happens as one gets progressively farther from node $a$ in Euclidean terms.
For this reason, the plots for $P_\delta(d)$ do not constitute a probability
distribution for any fixed value of $d$.

\begin{figure}[p]
\centering
\begin{tabular}{c}
\includegraphics[scale=0.65]{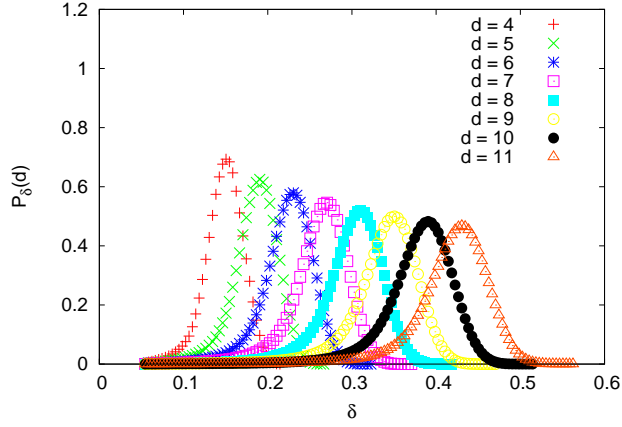}\\
{\small (a)}\\
\includegraphics[scale=0.65]{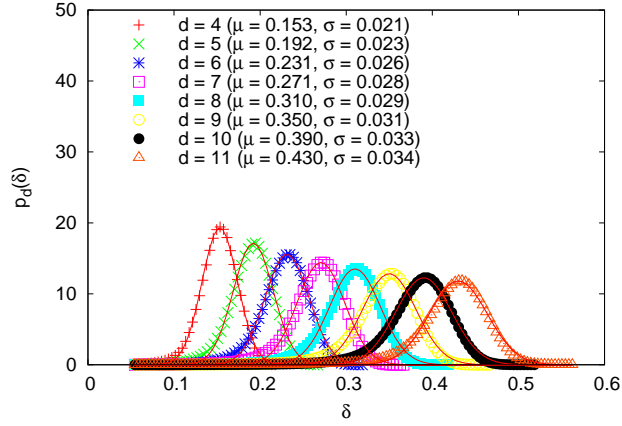}\\
{\small (b)}
\end{tabular}
\caption{$P_\delta(d)$ and $p_d(\delta)$ for $d>3$, with $z=3\pi$ (a, b) and
$z=5\pi$ (c, d). Solid lines give the Gaussians that best fit some of the
$p_d(\delta)$ data, each of mean $\mu$ and standard deviation $\sigma$ as
indicated.}
\label{fig:d>3}
\end{figure}

\addtocounter{figure}{-1}
\begin{figure}[p]
\centering
\begin{tabular}{c}
\includegraphics[scale=0.65]{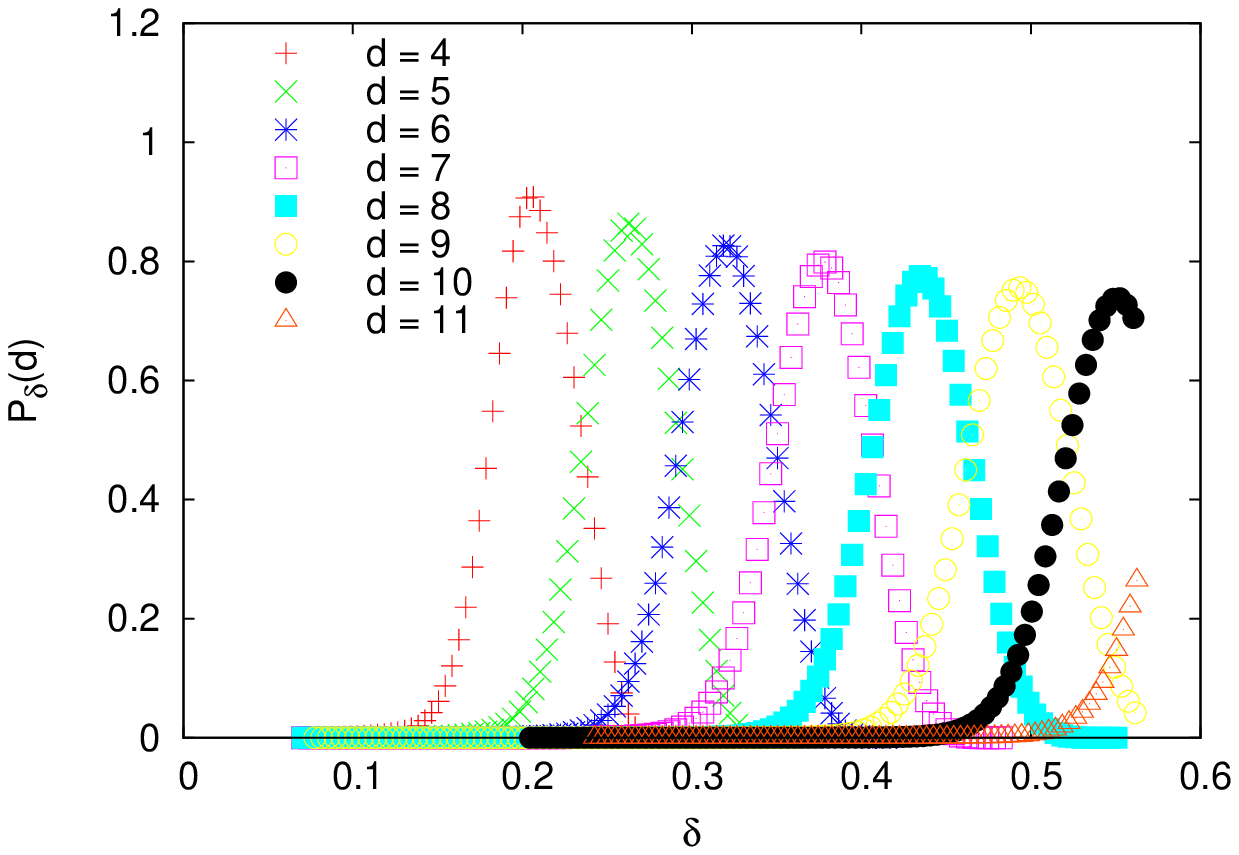}\\
{\small (c)}\\
\includegraphics[scale=0.65]{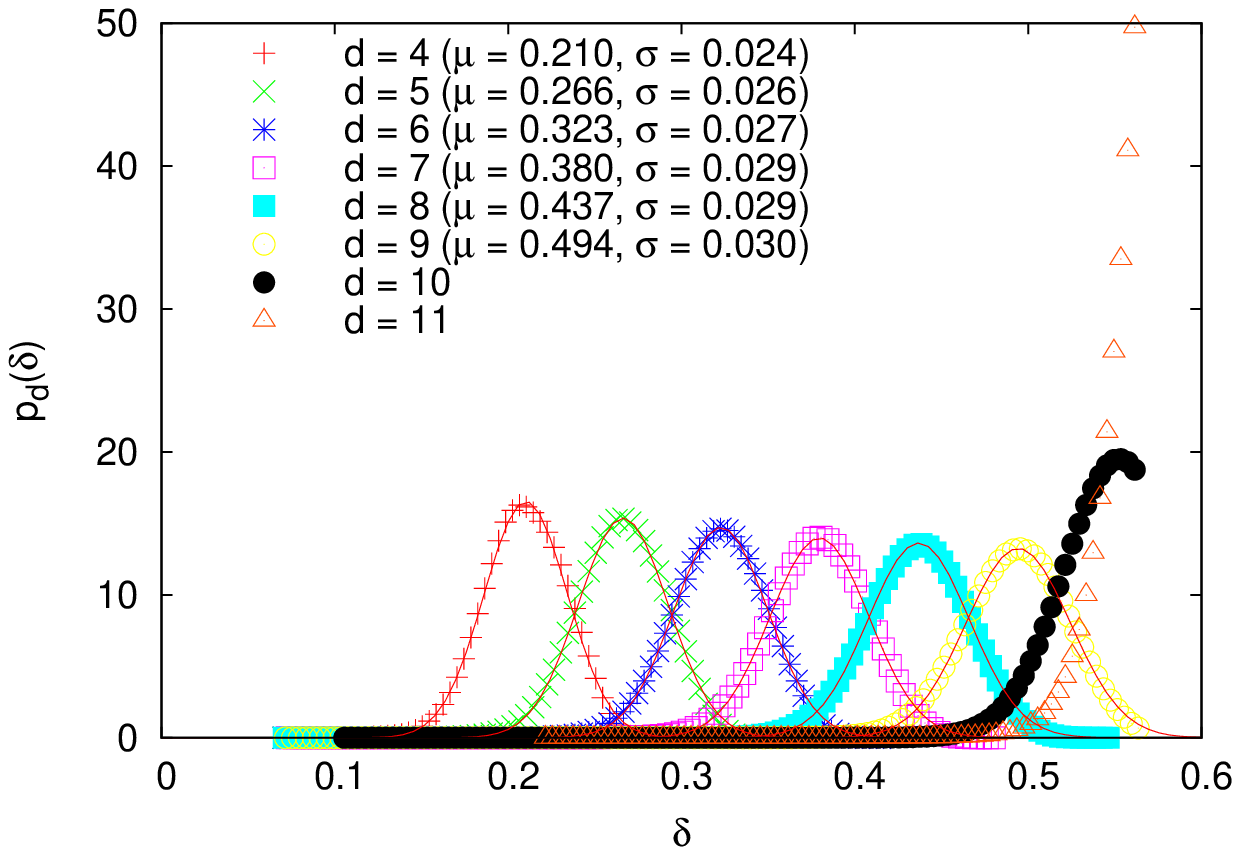}\\
{\small (d)}
\end{tabular}
\caption{Continued.}
\end{figure}

\section{Discussion and conclusion}

The results summarized in Figures~\ref{fig:d=1} through \ref{fig:d=3} reveal
excellent agreement between the analytic predictions we derived in Sections~3
through 5 and our simulation data. This holds not only for the simple cases of
$d=1$ and $d=2$, but also for the considerably more complex cases of
$P'_\delta(3)$ and $P_\delta(3)$. The latter, in particular, depends on the
empirically determined $n'$. In this respect, it is clear from
Figure~\ref{fig:d=3} that, even though $n'$ could have been calculated for a
greater assortment of $\delta$ values, doing it exclusively for $\delta=2R$
seems to have been sufficient.

Figure~\ref{fig:d>3} contemplates some of the $d>3$ cases, for which we derived
no analytic predictions. The values of $d$ that the figure covers in parts (a,
b) and (c, d), respectively for $z=3\pi$ and $z=5\pi$, are $4,\ldots,11$. Of
these, $d=11$ for $z=5\pi$ in part (d) typifies what happens for larger values
of $d$ as well (omitted for clarity), viz.\ probability densities sharply
concentrated at the border of the radius-$\sqrt{1/\pi}$ circle centered at node
$a$. Note that the same also occurs for $z=3\pi$, but owing to the smaller $R$
it only happens for larger values of $d$ (omitted from part (b), again for
clarity).

For $4\le d\le 11$ with $z=3\pi$, and $4\le d\le 9$ with $z=5\pi$,
Figures~\ref{fig:d>3}(b) and (d) also display Gaussian approximations of
$p_d(\delta)$. Parts (a) and (c) of the figure, in turn, contain the
corresponding simulation data only, and we remark that the absence of some
approximation computed from the Gaussians of part (b) or (d) is not a matter of
difficulty of principle. In fact, the counterpart of Equation~(\ref{eq:pfromP}),
obtained also from Bayes' theorem and such that
\begin{equation}
P_\delta(d)
=\frac{p_d(\delta)P(d)}{p(\delta)}
=\frac{p_d(\delta)P(d)}{\sum_{s=1}^{n-1}p_s(\delta)P(s)},
\end{equation}
can in principle be used with either those Gaussians or the concentrated
densities in place of $p_s(\delta)$ as appropriate for each $s$. What prevents
this, however, is that we lack a characterization of $P(s)$ that is not based on
simulation data only.

Still in regard to Figure~\ref{fig:d>3}, one possible interpretation of the good
fit by Gaussians of the simulation data for $p_d(\delta)$ comes from resorting
to the central limit theorem in its classical form \cite{trivedi}. In order to
do this, we view $\delta$ as valuing the random variable representing the
average Euclidean distance to node $a$ of all nodes that are $d$ edges apart
from $a$. The emergence of $p_d(\delta)$ as a Gaussian for $d>3$ (provided $d$
is small enough that the circle's border is not influential) may then indicate
that, for each value of $d$, the Euclidean distances of those nodes to node $a$
are independent, identically distributed random variables. While we know that
this does not hold for the smaller values of $d$ as a consequence of the
uniformly random positioning of the nodes in the circle (smaller Euclidean
distances to $a$ are less likely to occur for the same value of $d$), it would
appear that it begins to hold as $d$ is increased.

To summarize, we have considered a network of sensors placed uniformly at random
in a two-dimensional region and, for its representation as a random geometric
graph, have studied two distance-related distributions. One of them is the
probability distribution of distances between two randomly chosen nodes,
conditioned on the Euclidean distance between them. The other is the probability
density function associated with the Euclidean distance between two randomly
chosen nodes, given the distance between them. We have provided analytical
characterizations whenever possible, in the simplest cases as closed-form
expressions, and have also validated these predictions through simulations.

While further work related to additional analytical characterizations is worth
undertaking, as is the investigation of the three-dimensional case, we find that
the most promising tracks for future investigation are those that relate to
applications. In Section~1 we illustrated this possibility in the context of
sensor localization, for which it seems that understanding the distance-related
distributions we have studied has the potential to help in the discovery of
better distributed algorithms. Whether there will be success on this front
remains to be seen, as well as whether other applications will be found with the
potential to benefit from the results we have presented.

\subsection*{Acknowledgments}

The authors acknowledge partial support from CNPq, CAPES, and a FAPERJ BBP
grant.

\bibliography{distprob}
\bibliographystyle{plain}

\end{document}

%% file: tab1.tex
\begin{tabular}{|c|c|c|c|c|c|c|}
\hline
$\delta^-$
 &$\delta^+$
 &$x_k^-$
 &$x_k^+$
 &$y_k^-(x_k)$
 &$y_k^+(x_k)$
 &Fig.\\
\hline\hline
$R$
 &$R\sqrt{3}$
 &$x_A$
 &$x_B$
 &$y_D-\sqrt{R^2-(x_k-x_D)^2}$
 &$y_D+\sqrt{R^2-(x_k-x_D)^2}$
 &\ref{fig:Kdelta}(a)\\
&&$x_B$
 &$0$
 &$y_D-\sqrt{R^2-(x_k-x_D)^2}$
 &$\sqrt{R^2-x_k^2}$
 &\\
&&$0$
 &$x_C$
 &$0$
 &$\sqrt{R^2-x_k^2}$
 &\\
&&$x_C$
 &$x_D$
 &$\sqrt{R^2-(x_k-\delta)^2}$
 &$\sqrt{R^2-x_k^2}$
 &\\
\hline
$R\sqrt{3}$
 &$2R$
 &$x_A$
 &$0$
 &$y_D-\sqrt{R^2-(x_k-x_D)^2}$
 &$y_D+\sqrt{R^2-(x_k-x_D)^2}$
 &\ref{fig:Kdelta}(b)\\
&&$0$
 &$x_B$
 &$0$
 &$\sqrt{4R^2-(x_k-\delta)^2}$
 &\\
&&$x_B$
 &$x_C$
 &$0$
 &$\sqrt{R^2-x_k^2}$
 &\\
&&$x_C$
 &$x_D$
 &$\sqrt{R^2-(x_k-\delta)^2}$
 &$\sqrt{R^2-x_k^2}$
 &\\
\hline
$2R$
 &$3R$
 &$x_A$
 &$x_B$
 &$0$
 &$\sqrt{4R^2-(x_k-\delta)^2}$
 &\ref{fig:Kdelta}(c)\\
&&$x_B$
 &$R$
 &$0$
 &$\sqrt{R^2-x_k^2}$
 &\\
\hline
\end{tabular}

%% file: tab2.tex
\begin{tabular}{|c|c|c|c|c|c|c|c|}
\hline
$\delta^-$
 &$\delta^+$
 &$x_A$
 &$x_B$
 &$x_C$
 &$x_D$
 &$y_D$
 &Fig.\\
\hline\hline
$R$
 &$R\sqrt{3}$
 &$\delta/2-R$
 &$\left(\delta-\sqrt{3(4R^2-\delta^2)}\,\right)/4$
 &$\delta-R$
 &$\delta/2$
 &$\sqrt{4R^2-\delta^2}/2$
 &\ref{fig:Kdelta}(a)\\
\hline
$R\sqrt{3}$
 &$2R$
 &$\delta/2-R$
 &$(\delta^2-3R^2)/2\delta$
 &$\delta-R$
 &$\delta/2$
 &$\sqrt{4R^2-\delta^2}/2$
 &\ref{fig:Kdelta}(b)\\
\hline
$2R$
 &$3R$
 &$\delta-2R$
 &$(\delta^2-3R^2)/2\delta$
 &&&&\ref{fig:Kdelta}(c)\\
\hline
\end{tabular}